\documentclass[conference]{worldcomp}

\usepackage[hmargin=.75in,vmargin=1in]{geometry}
\usepackage[american]{babel}
\usepackage[T1]{fontenc}
\usepackage{times}

\usepackage{graphicx,url,verbatim}
\usepackage[smaller]{acronym}
\usepackage[square,numbers]{natbib}
\usepackage{algorithm}
\usepackage{algpseudocode}
\usepackage{amssymb}
\usepackage{fancyhdr}

\usepackage{flushend}

\title{\bf Distributed Parallel D8 Up-Slope Area Calculation in Digital Elevation Models}
\author{
{\bfseries R. Barnes$^1$, C. Lehman$^2$, and D. Mulla$^3$}\\
$^1$Ecology, Evolution, \& Behavior, University of Minnesota, Minneapolis, MN, USA\\
$^2$College of Biological Sciences, University of Minnesota, Minneapolis, MN, USA\\
$^3$Soil, Water, \& Climate, University of Minnesota, Minneapolis, MN, USA\\
}

\graphicspath{{pics/}}

\acrodef{DEM}{digital elevation model}

\newcommand{\algofunc}[1]{\textsf{\textbf{{#1}}}}
\algblockdefx{MultiForAll}{EndMultiFor}{\textbf{for$_\parallel$ all }}{\textbf{end for$_\parallel$}}

\newcommand{\thispageheader}[2][R]{\expandafter\def\csname header#1\endcsname{#2}}

\begin{document}

\twocolumn[
  \begin{@twocolumnfalse}
    \noindent Cite as: Barnes, Lehman, Mulla. ``Distributed Parallel D8 Up-Slope Area Calculation in Digital Elevation Models". 2011 International Conference on Parallel and Distributed Processing Techniques and Applications. Las Vegas, NV. p. 833--838.

		{\let\newpage\relax\maketitle}
  \end{@twocolumnfalse}
]

\begin{abstract}
This paper presents a parallel algorithm for calculating the eight-directional (D8) up-slope contributing area in \acp{DEM}. In contrast with previous algorithms, which have potentially unbounded inter-node communications, the algorithm presented here realizes strict bounds on the number of inter-node communications. Those bounds in turn allow D8 attributes to be processed for arbitrarily large DEMs on hardware ranging from average desktops to supercomputers. The algorithm can use the OpenMP and MPI parallel computing models, either in combination or separately. It partitions the DEM between slave nodes, calculates an internal up-slope area by replacing information from other slaves with variables representing unknown quantities, passes the results on to a master node which combines all the slaves' data, and passes information back to each slave, which then computes its final result. In this way each slave's DEM partition is treated as a simple unit in the DEM as a whole and only two communications take place per node.
\end{abstract}

\vspace{1em}
\noindent\textbf{Keywords:}
 {\small  D8, up-slope area, digital elevation model, distributed, parallel, flow accumulation} %%%% Max of 6

\section{Introduction}
\acp{DEM} are data structures, usually a rectangular array of floating-point or integer values, representing terrain elevation above some common base level, generally measured via remote sensing techniques or LIDAR.

\acp{DEM} are used extensively to model hydrologic processes and properties including soil moisture (based on catchment area), terrain instability (based on slope and catchment area), erosion (based on slope), and stream power (based on slope and catchment area) \citep{Wallis2009}.

Underlying the aforementioned calculations is a flow function responsible for determining what proportion of each cell's flow each neighbouring cell will receive. Perhaps the most widely used function is the D$_8$ function introduced by \citet{Ocallaghan1984}. This function directs the entirety of each cell's flow to the lowest of its eight surrounding neighbours. This implies that flows \emph{combine but never disperse}: a property we take advantage of. In contrast is the D$_\infty$ method introduced by \citet{Tarboton1997}, which calculates an angle of steepest descent based on adjacent pairs of neighbouring cells and directs flow to one or both neighbours along that path.

The accuracy of \ac{DEM}-based calculations is related to the \ac{DEM}'s resolution. These have gone from thirty-plus meter resolution in the recent past to the sub-meter resolutions becoming available today. Increasing resolution has led to increased data sizes: current data sets are on the order of gigabytes and increasing. While computer processing and memory performance have increased appreciably during this time, legacy equipment and algorithms suited to manipulating smaller \ac{DEM}s with coarser resolutions make processing these improved data sources costly, if not impossible.

\citet{Wallis2009} present one solution for calculating ``up-slope area" based on previous work by \citet{mark1988} and \citet{Ocallaghan1984}: a parallel algorithm suitable for both D$_8$ and D$_\infty$ calculations for use in environments where communication is inexpensive using a queue-based up-slope area function. Although their algorithm, as published, assumes shared memory, this is not a strict requirement.

The algorithm presented here applies only to D$_8$, is suitable for environments where inter-node communication is expensive relative processing, makes efficient use of multi-processor nodes, and optimises the queue method presented by \citet{Wallis2009}.

\section{Up-Slope Area}
Up-slope area $A$ is defined physically for each point $p$ in a watershed as the set of all points $P$ whose flow, if we were to put a liquid in them, would eventually pass through $p$. Mathematically, this is defined by the recursion relation:
\begin{equation}
A(p)=1+\sum_{i=n(p,P)} A(i)
\end{equation}
Where $n(p,P)$ defines the points neighbouring $p$, given $P$. In practice, points are generally represented by cells of some sort, which may or may not be of equal area and may or may not be weighted equally in the calculation. In the case of most \acp{DEM} and this paper, the points are represented by square cells with an area and weight of one.

\section{The Algorithm}
It is assumed that the \ac{DEM} has been preprocessed to remove pits and flats. Pits are cells with no lower neighours or groups of cells from which there is no outlet. In the case of a single cell, the cell is generally raised to the level of its lowest neighbor; in the case of groups, an outlet is usually drilled. Flats are cells with one or more neighbours of equal elevation, these my be resolved by making slight elevation modifications within the flat; \citet{Garbrecht1997} present one approach.
%TODO and TODO present approaches to this. Pit filling

Following this, each of $S$ slave nodes reads in an equal number of horizontal rows of the \ac{DEM}, with any extra rows allocated to the final slave. In addition to its allotted rows, each slave also reads in two extra rows above and below its strip. This permits the slave to determine flow direction for the cells at the edge of its allotted strip and for the cells bordering it. Henceforth, cells at the edge of a slave's allotted strip are called \emph{edge cells} and are found in either the $TopRow$ or the $BottomRow$ of the strip.

In this algorithm, the D$_8$ flow direction of a cell is specified as being towards the neighbour with the steepest slope relative that cell using the Euclidean distance between centers of the two cells. We call the net flow field $F$.

%If the \ac{DEM} is $H$ cells high and $W$ cells wide, then each slave $s$ will read in the horizontal strip of rows $\left[\max(0,p\lfloor\frac{H}{S}\rfloor-2) , \min(H,(p+1)\lfloor\frac{H}{S}\rfloor+2\right]$. The two extra rows above and below each slave's nominal\footnote{todo: word choice} strip permit 

Throughout this paper procedures marked with a parallel subscript (``$_\parallel$") may be safely invoked on multiple processors whereas loops marked with a parallel subscript be safely partitioned between processors. Any commands which may lead to race conditions are marked as atomic.

Using the flow field $F$, each slave finds the dependencies of all its cells using Algorithm \ref{alg:finddeps}. A cell $c$ is a dependent of a neighbour $n$ if $n$'s flow is directed into it; thus, a cell may have 0--8 dependencies, inclusive. If a cell has no dependencies, it is pushed on to the back of a double-ended queue $Q$; otherwise, the number of dependencies is stored in an array $D$.

\begin{algorithm}
\caption{Slaves calculate dependencies}
\label{alg:finddeps}
\begin{algorithmic}[1]
\Procedure{FindDependencies}{}
\Require $F,D,Q$
	\MultiForAll{$c$ in $F$}\label{alg:finddeps-multifor}
		\ForAll{$n$ inputs to $c$}
			\State $D(c)\gets D(c)+1$
		\EndFor
		\If{$D(c)=0$}
			\State \Call{Atomic}{push $c$ onto back of $Q$}\label{alg:finddeps-atomic}
		\EndIf
	\EndMultiFor
\EndProcedure
\end{algorithmic}
\end{algorithm}

The purpose of the double-ended queue is to minimize contention in multi-processor environments by decoupling the queue's push and pop functions. The \algofunc{for$_\parallel$} loop on Line \ref{alg:finddeps-multifor} may be safely run in parallel because none of the data being read is altered by the function and only one processor will be writing to any given memory location. The queue-push on Line \ref{alg:finddeps-atomic} must be done atomically for the algorithm to run safely. It is possible to implement Algorithm \ref{alg:finddeps} such that there is no contention by having each processor build its own queue and then merging these just after the function completes.

Following Algorithm \ref{alg:finddeps}, each slave performs calculates its Internal Up-slope Area $A$ using Algorithm \ref{alg:internalupslope}. This algorithm begins with the local maxima of the \ac{DEM}---those cells added to $Q$ by Algorithm \ref{alg:finddeps}---which have an up-slope area of one (only themselves) and decrements the dependency counter of the cells they flow into. When a cell's dependency counter reaches zero, it is added to $Q$. Thus, as the up-slope area of higher-elevation cells becomes known, it is possible to calculate that of lower-elevation cells.

\begin{algorithm}
\caption{Slaves calculate internal up-slope area}
\label{alg:internalupslope}
\begin{algorithmic}[1]
\Procedure{InternalUpslope$_\parallel$}{$c$}
\Require $F, D, Q, A$
	\If{$c$ was not specified}
		\State \Call{Atomic}{$c\gets$ front of $Q$}\label{alg:internalupslope-atomic}
		\If{$c$ was not set}
			\State \Return
		\EndIf
	\EndIf
	\Statex
	\State $A(c)\gets 1$ \label{alg:internalupslope-omit1}
	\ForAll{$n$ inputs to $c$}
		\State $A(c)\gets A(c)+A(n)$
	\EndFor
	\Statex
	\State $n\gets$ downslope neighbor of $c$
	\If{$n$ exists}
		\State $D(n)\gets D(n)-1$
		\If{$D(n)\neq0$}
			\State $n\gets$\Call{NULL}{}
		\EndIf
	\EndIf
	\State \Return \Call{InternalUpslope}{$n$} \label{alg:internalupslope-recur}
\EndProcedure
\end{algorithmic}
\end{algorithm}

Algorithm \ref{alg:internalupslope} is defined using tail recursion and, to avoid excessive stack sizes, should be implemented appropriately. Since $Q$ cannot be equitably divided among processors beforehand, it is necessary for each processor to atomically pop cells from it when necessary. However, the depth-first search embodied by the recursion on Line \ref{alg:internalupslope-recur} reduces contention on $Q$.

\begin{figure}
\center
\includegraphics[width=\linewidth]{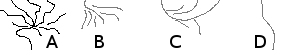}
\caption{Conceptual image of dependencies and resolutions}
\label{fig:externalconceptual}
\end{figure}

At this point, information from other slaves is required to resolve the remaining dependencies. Figure \ref{fig:externalconceptual} depicts this conceptually. \textbf{A} represents a flow path originating in the current slave's portion of the \ac{DEM}: all its dependencies can be satisfied with the information available to the slave, but these results must be communicated to neighbouring slaves.

If we assume that flow is generally directed to the lower-right, then \textbf{B} represents an up-slope area originating in a different slave and terminating in this one. In such a case a single communication between the slaves provides sufficient information to compute up-slope area.

\textbf{C} represents a similar situation, but two communications are necessary because the flow path loops back up: one to resolve this slave's dependencies and another to resolve its neighbour's.

\textbf{D} represents a situation in which two communications are necessary to pass information from one neighbouring slave to another through the current slave.

If \textbf{C} and \textbf{D} link through the upper neighbouring slave, then four communications would be necessary. Or, if flow is directed to the upper-left, and \textbf{B}, \textbf{C}, and \textbf{D} are linked, then five communications would be necessary. We denote the average number of communications a slave will need to make as its ``winding factor" $\phi$. In situations where communication is cheap, $\phi$ is unimportant and algorithms such as that presented by \citet{Wallis2009} are effective. In situations where communication is expensive, a different algorithm, such as that presented here, is necessary.

If each slave denotes external inputs from other slaves by variables, it is possible to continue the calculation with minimally inter-nod communication.

We call edge cells with unresolved dependencies in other slaves' \acp{DEM} \emph{receivers}. They receive information from at most 3 of the other slave's edge cells. Cells upon which another slave's \ac{DEM} depends are either \emph{givers} or \emph{joiners}. A giver cell's upslope-area has already been calculated by Algorithm \ref{alg:internalupslope}; a joiner is the dependent of one or more receiver cells.

It is a property of the D$_8$ flow function that flows join but never disperse and, therefore, that every joiner is a dependent of \emph{at least} one receiver while each receiver ultimately feeds into \emph{at most} one joiner. This property does not hold for the D$_\infty$ flow function, preventing its use in this context.

Algorithm \ref{alg:satisfyreceivers} presents one possible way of preparing the slave to perform the External Up-slope Area calculation. The original dependency grid is saved on Line \ref{alg:satisfyreceivers-save} to be used later in calculating the true up-slope area. Once this is done, the external dependencies of all edge cells are removed and those cells which are ``satisfied" are added to $Q$. The cell is marked as a receiver by assigning it a special variable name on Line \ref{alg:satisfyreceivers-receiver}.

$Cell_V$, first mentioned on Line \ref{alg:satisfyreceivers-receiver}, is a map between cells $c$ and variable names, which are represented as globally-unique numbers with $-1$ acting as a special value used to denote receivers. Algorithm \ref{alg:satisfyreceivers} marks all cells as receivers; Algorithm \ref{alg:externalupslope} will remark those which are not.

\begin{algorithm}
\caption{Slaves prepare to calculate external up-slope area}
\label{alg:satisfyreceivers}
\begin{algorithmic}[1]
\Procedure{SatisfyReceivers}{}
\Require $F, D, D_O, Cell_V, Q$
	\State $D_O\gets D$ \label{alg:satisfyreceivers-save}
	\MultiForAll{$c$ in $TopRow$}
		\State $V\gets$\Call{NewVariable}{}
		\State $Cell_V(c)\gets-1$ \label{alg:satisfyreceivers-receiver}
		\ForAll{$n$ above-inputs to $c$}
			\State $D(c)\gets D(c)-1$
		\EndFor
		\If{$D(c)=0$}
			\State \Call{Atomic}{push $c$ onto back of $Q$}
		\EndIf
	\EndMultiFor
	\algrenewcommand{\algorithmiccomment}[1]{$\triangleright$ #1}
	\State \Comment{Analogous code for $BottomRow$}
\EndProcedure
\end{algorithmic}
\end{algorithm}

Algorithm \ref{alg:externalupslope} calculates External Up-slope Area, keeping track of how cells depend on receivers. When a cell is popped off $Q$, it is associated with a new variable name $V$ (Line \ref{alg:externalupslope-newvar1}) which is unique across all slaves and used in future iterations to keep track of the flow path's origin using the multimap\footnote{A hash table which associates the same key with multiple values. In this paper empty or nonexistent keys in maps and multimaps always return NULL when their values are cell identifiers and 0 when their values are used mathematically.} $Origin$ (Line \ref{alg:externalupslope-newvar2}). Since only edge cells are in $Q$, new variables are only formed at the edge cells.

After incrementing a cell $c$'s up-slope area, the algorithm attempts to follow the flow path from $c$ to its neighbour $n$ (Line \ref{alg:externalupslope-follow}). If there is a neighbour and its dependencies are satisfied, the algorithm recurses, maintaining knowledge of which edge cell its flow path originated at using $V$ (Line \ref{alg:externalupslope-satisfied}). If $n$'s dependencies are not satisfied, the algorithm uses the map $Cell_V$ to inform a future iteration of the existence of the flow path it's about to abandon (Line \ref{alg:externalupslope-remember}).

On Lines \ref{alg:externalupslope-paban1}--\ref{alg:externalupslope-paban2}, after the cell's up-slope area has been incremented, the algorithm inspects the present cell and and cells flowing into it to see if it or they are part of a previously abandoned flow path. If so, it merges that flow path with the present one and continues.

\begin{algorithm}
\caption{Slaves calculate external up-slope area}
\label{alg:externalupslope}
\begin{algorithmic}[1]
%Todo: Needs to have Cell_V and Origin variables included. Consider moving globals out of call list
\Procedure{ExternalUpslope$_\parallel$}{$V, c$}
\Require $F, D, Q, A, Cell_V, Origin$
	\If{$c$ was not specified}
		\State \Call{Atomic}{$c\gets$ front of $Q$}
		\If{$c$ was not set}
			\State \Return \label{alg:externalupslope-exit}
		\EndIf
		\State $V\gets$\Call{NewVariable}{} \label{alg:externalupslope-newvar1}
		\State $Origin(V)\gets c$ \label{alg:externalupslope-newvar2}
	\EndIf
	\Statex
	\State $A(c)\gets 1$
	\ForAll{$n$ inputs to $c$}
		\State $A(c)\gets A(c)+A(n)$
	\EndFor
	\Statex
	\ForAll{$n$ inputs to $c$ and $c$ itself}	\label{alg:externalupslope-paban1}
		\State append $Origin(Cell_V(n))$ to $Origin(V)$
		\State erase $Origin(Cell_V(n))$
		\State erase $Cell_V(n)$
	\EndFor \label{alg:externalupslope-paban2}
	\Statex
	\State $n\gets$ downslope neighbor of $c$ \label{alg:externalupslope-follow}
	\If{$n$ exists}
		\State $D(n)\gets D(n)-1$
		\If{$D(n)=0$}
			\State \Return \Call{ExternalUpslope}{$V, n$} \label{alg:externalupslope-satisfied}
		\EndIf
	\EndIf
	\State $Cell_V(c)=V$ \label{alg:externalupslope-remember}
	\State \Return \Call{ExternalUpslope}{$-, -$}
\EndProcedure
\end{algorithmic}
\end{algorithm}

Ultimately, the algorithm reaches a point where no down-slope neighbour exists and there are no more cells in $Q$. It abandons the flow path (Line \ref{alg:externalupslope-remember}), recurses, and exits (Line \ref{alg:externalupslope-exit}). The result of this algorithm is a map $Cell_V$ indexed on the joiner cells. These cells are part of the border of the current slave and the edge of the adjacent slave---they are the adjacent slave's receivers. So $Cell_V$, coupled with $Origin$ link one slave's receivers to another's.

Each slave now sends $Cell_V$, $Origin$, and the up-slope area of its bordering cells to the master node.

\begin{figure}
\center
\includegraphics[width=\linewidth]{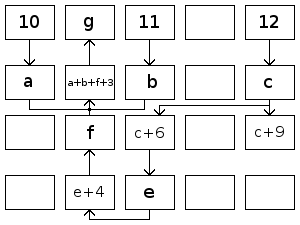}
\caption{Conceptual image of master node's data}
\label{fig:gridconceptual}
\end{figure}

Conceptually, the situation is exemplified by Figure \ref{fig:gridconceptual}. Data from different slaves is separated by wide vertical gaps; it is important to remember that the cells bordering these gaps, though they appear in different slaves are, in fact, the same in the \ac{DEM}. Within a slave's data, receivers and joiners form sometimes-complicated linkages while between slaves the linkages are simple; from the master's perspective, the slaves combine to form a directed acyclic graph.

In Figure \ref{fig:gridconceptual}, receivers are marked by variables. These are passed along to joiners, accumulating up-slope area along the way. Therefore, the joiners are represented by the sum of their associated receivers' variables and the upslope area connecting them to each receiver. Givers are represented by a pure number: this is a true up-slope area and a starting point for the next calculation.

Algorithm \ref{alg:masterprep} prepares the slaves' data for processing by locating giver cells and determining how many dependencies each joiner has. Since the map keys are globally unique, all the slaves' $Cell_V$s are combined; likewise, the $Origin$s. The function \emph{Invert} on Line \ref{alg:masterprep-invert} turns $Origin$'s keys into values and vice versa. The result is a simple map serving the same purpose as the flow field array $F$ of Algorithm \ref{alg:internalupslope}. The dependency array holds: 1 for receivers since they only depend on data from cells in adjacent slaves; -1 for givers so that they are not later mistaken for receivers; and a positive number, 1 or more, for joiners.

\begin{algorithm}
\caption{Master node prepares received data}
\label{alg:masterprep}
\begin{algorithmic}[1]
\Procedure{MasterPrep}{}
\Require $D, Q, Cell_V, Origin$
	\MultiForAll{Slaves $s$}
		\ForAll{$c$ in $(TopRow,BottomRow)$}
			\If{$Cell_V(c)$ is undefined} \Comment{Giver}
				\State \Call{Atomic}{push $c$ onto back of $Q$}
				\State $D(c)\gets-1$
			\ElsIf{$Cell_V(c)=-1$}
				\Statex \Comment{Receiver}
				\State $D(c)\gets1$
			\Else \Comment{Joiner}
				\State $D(c)\gets$\Call{Length}{$Origin(Cell_V(C))$}
			\EndIf
		\EndFor
	\EndMultiFor
	\State $Destination\gets$\Call{Invert}{$Origin$} \label{alg:masterprep-invert}
\EndProcedure
\end{algorithmic}
\end{algorithm}

Algorithm \ref{alg:masterupslope} solves the system of equations presented by the slaves using the same methodology as in Algorithms \ref{alg:internalupslope} and \ref{alg:externalupslope}. Lines \ref{alg:masterupslope-dest1}--\ref{alg:masterupslope-dest2} identify the cell's destination. If one has not been explicitly declared the cell must either be a giver or a joiner and its destination is implied as being to the slave above/below it, provided that cell is a receiver. Figure \ref{fig:gridconceptual} depicts one situation wherein this would arise. The destination of the cell in the bottom right labeled ``$c+9$" is implied to be directly below it; however, the two are unconnected. If the destination is valid, the algorithm continues in the usual fashion.

\begin{algorithm}
\caption{Master node calculates up-slope area}
\label{alg:masterupslope}
\begin{algorithmic}[1]
\Procedure{MasterUpslope$_\parallel$}{$c$}
\Require $A, D, Q, Cell_V, Destination$
	\If{$c$ was not specified}
		\State \Call{Atomic}{$c\gets$ front of $Q$}
		\If{$c$ was not set}
			\State \Return
		\EndIf
	\EndIf
	\Statex
	\State $n\gets Destination(c)$\label{alg:masterupslope-dest1}
	\If{$n$ is undefined}
		\If{$c$ is a top cell}
			\State $n\gets$cell above $c$
		\Else
			\State $n\gets$cell below $c$
		\EndIf
		\If{$Cell_V(n)\neq-1$}\Comment{Not A Receiver}
			\State $n\gets$\Call{NULL}{}
		\EndIf
	\EndIf\label{alg:masterupslope-dest2}
	\If{$n$ is undefined}
		\State \Return \Call{MasterUpslope}{$-$}
	\EndIf
	\Statex
	\State $A(n)\gets A(n)+A(c)$
	\State $D(n)\gets D(n)-1$
	\If{$D(n)\neq0$}
		\State $n\gets$\Call{NULL}{}
	\EndIf
	\State \Return \Call{MasterUpslope}{$n$}
\EndProcedure
\end{algorithmic}
\end{algorithm}

\begin{algorithm}
\caption{Slaves prepare incoming data}
\label{alg:slaves_prepare_incoming}
\begin{algorithmic}[1]
\Procedure{PrepFinaliseInternal}{}
\Require $F, D, D_O, Cell_V, Q$
	\State $D\gets D_O$ \label{alg:slaves_prepare_incoming-save}
	\MultiForAll{$c$ in $TopRow$}
		\ForAll{$n$ above-inputs to $c$}
			\State $D(c)\gets D(c)-1$
		\EndFor
		\If{$D(c)=0$}
			\State \Call{Atomic}{push $c$ onto back of $Q$}
			\State $Area_D(c)\gets A_{incoming}(c)$ \label{alg:slaves_prepare_incoming-area_incoming}
		\EndIf
	\EndMultiFor
	\algrenewcommand{\algorithmiccomment}[1]{$\triangleright$ #1}
	\State \Comment{Analogous code for $BottomRow$}
\EndProcedure
\end{algorithmic}
\end{algorithm}

Once Algorithm \ref{alg:masterupslope} is completed, the area of each slaves' top and bottom rows are returned. Since Algorithm \ref{alg:externalupslope} set the areas of the slaves' receivers to one and calculated up-slope area that, we simply have to add each receivers' incoming area to all its dependents. Line \ref{alg:slaves_prepare_incoming-area_incoming} of Algorithm \ref{alg:slaves_prepare_incoming} enables this by setting up a map between receivers and their incoming variables. This map will later be used to keep track of which incoming areas belong to which flow path. Finally, the slaves run Algorithm \ref{alg:finalinternal}.

Algorithm \ref{alg:finalinternal} is similar to Algorithm \ref{alg:externalupslope} insofar as it propagates variables forward. However, rather than propagating variables and combining flow paths, only the incoming areas are propagated forward.

\begin{algorithm}
\caption{Slaves finalise internal up-slope areas}
\label{alg:finalinternal}
\begin{algorithmic}[1]
%Todo: Needs to have Cell_V and Origin variables included. Consider moving globals out of call list
\Procedure{FinaliseInternal$_\parallel$}{$S, c$}
\Require $F, D, Q, A$
	\If{$c$ was not specified}
		\State \Call{Atomic}{$c\gets$ front of $Q$}
		\If{$c$ was not set}
			\State \Return
		\EndIf
		\State $S\gets Area_D(c)$
	\EndIf
	\Statex
	\ForAll{$n$ inputs to $c$}
		\State $S\gets S+Area_D(c)$
		\State erase $Area_D(c)$
	\EndFor
	\State $A(c)\gets A(c)+S$
	\Statex
	\State $n\gets$ downslope neighbor of $c$
	\If{$n$ exists}
		\State $D(n)\gets D(n)-1$
		\If{$D(n)=0$}
			\State \Return \Call{FinaliseInternal}{$S, n$}
		\EndIf
	\EndIf
	\State $Area_D(c)=S$
	\State \Return \Call{FinaliseInternal}{$-, -$}
\EndProcedure
\end{algorithmic}
\end{algorithm}

%\section{Complexity and Efficiency Comparison}

%\section{Implementation Notes}

\section{Conclusions}
The algorithm presented here makes efficient use of multi-processor nodes and is well-suited to environments where communication is expensive and must be kept to a minimum. Each slave communicates with a master node only once and the master node communicates with each slave only once. In the case of a single node performing the calculations, it would be possible to store intermediate results to disk or layer this algorithm, allowing the node to process arbitrarily large \acp{DEM} dependent only on disk space.

\bibliographystyle{plainnat}
%\bibliography{library,others}
%\bibliographystyle{plainnat}
{\footnotesize
\bibliography{library,others}}

\end{document}